\documentclass[
    a4paper,
    notitlepage,
    pre,
    twocolumn,
    superscriptaddress,
    floatfix
]{revtex4-1}
\usepackage{amsmath,amssymb}
\usepackage{lipsum}
\usepackage{bm}
\usepackage{graphicx}
\usepackage{graphics}
\usepackage{amsmath}
\usepackage{array}
\usepackage[caption=false]{subfig}
\usepackage{xcolor}
\usepackage{subfig}
\usepackage{hyperref}
\usepackage[capitalise]{cleveref}
\usepackage{textcomp}

\usepackage{epstopdf}
\usepackage{cases}
\usepackage{amssymb}
\usepackage{multirow}
\usepackage{siunitx}
\usepackage{cases}
\usepackage{tabularx}
\usepackage[top=1.1in, bottom=1.1in, left=1.0in, right=1.0in]{geometry}
\usepackage[utf8]{inputenc}
\usepackage{tikz}
\usepackage{comment}
\usepackage{lineno,hyperref}
\modulolinenumbers[5]
\usepackage{booktabs}

\begin{document}

\title{Non-monotonic heat dissipation in close-packed quasi-2D and 3D hotspot system}
\author{Chuang Zhang}
\email{zhangc33@sustech.edu.cn}
\affiliation{Department of Mechanics and Aerospace Engineering, Southern University of Science and Technology, Shenzhen 518055, China}
\author{Lei Wu}
\email{Corresponding author: wul@sustech.edu.cn}
\affiliation{Department of Mechanics and Aerospace Engineering, Southern University of Science and Technology, Shenzhen 518055, China}
\date{\today}

\begin{abstract}

Transient heat dissipation in close-packed quasi-2D nanoline and 3D nanocuboid hotspot systems is studied based on phonon Boltzmann transport equation.
It is found that, counter-intuitively, the heat dissipation efficiency is not a monotonic function of the distance between adjacent nanoscale heat sources:
{\color{red}{the heat dissipation efficiency reaches the highest value when this distance is comparable to the phonon mean free path.}} This is due to the competition of two thermal transport processes: quasiballistic transport when phonons escape from the nanoscale heat source and the scattering among phonons originating from adjacent nanoscale heat source.

\end{abstract}

\maketitle

\section{Introduction}

With the fast development of micro- and nanotechnologies~\cite{cahill2003nanoscale,cahill2014nanoscale,moore_emerging_2014} and the drastically reduced size of electronic devices~\cite{moore_emerging_2014,RevModPhys.90.041002}, the Moore's law is reaching its limit.
Besides, the increase of power density intensifies hotspot issues and increases the demand for heat dissipation.
The heat dissipation problem at the micro/nano scale has become one of the key bottlenecks restricting the further development of the microelectronics industry.
Hence, it is important to understand the thermal transport mechanisms in microelectronic devices~\cite{warzoha_applications_2021,moore_emerging_2014,yue_advances_2021} to realize optimal waste heat removal and improve device performance and reliability.

At the micro/nano-scale, the Fourier law of thermal conduction becomes invalid and the non-Fourier phonon transport is summarized into four major categories~\cite{cahill2003nanoscale,cahill2014nanoscale,ZHANG2020,chen_non-fourier_2021,RevModPhys.90.041002}.
The first is the ballistic phonon transport~\cite{MajumdarA93Film,zhang_discrete_2019,xu_raman-based_2020}, which happens when the systems characteristic length/time is much smaller/shorter than the phonon mean free path~\cite{ju1999phonon,hsiao_observation_2013,chang_breakdown_2008,xu_length-dependent_2014}/relaxation time~\cite{beardo_observation_2021,GuoZl16DUGKS,LUO2017970}.
The second arises from small-scale heat sources~\cite{PhysRevB.38.1963,chen1996,ballisticsiliconapl2001,siemens_quasi-ballistic_2010,PhysRevLett.107.095901,hu_spectral_2015,chuang2021graded}.
When a hotspot with small size is added in a bulk material, if the phonon mean free path is much larger than the size of hotspot, phonons emitted from the hotspot do not suffer sufficient phonon-phonon scattering near the hotspot region so that quasiballistic phonon transport occurs even if there is no boundary or interface scattering inside the systems~\cite{ballisticsiliconapl2001,siemens_quasi-ballistic_2010,hu_spectral_2015}.
The third is the coherent phonon transport~\cite{luckyanova_phonon_2018,PhysRevX.10.021050,ma2019b}, which appears when the systems characteristic length is comparable to the phonon wavelength.
The fourth is the hydrodynamic phonon transport, which requires that the momentum-conserved normal scattering to be much more extensive than the boundary scattering and the boundary scattering is much more sufficient than the momentum-destroying resistive scattering~\cite{lee_hydrodynamic_2015,cepellotti_phonon_2015,huberman_observation_2019}.
{\color{red}{So far, the phonon hydrodynamics phenomena have only been experimentally measured in a few three-dimensional (3D) materials (e.g., graphite, NaF) at low temperatures~\cite{PhysRevLett.16.789,PhysRevLett_secondNaF,huberman_observation_2019}.}}

Except above situations, recent studies have revealed the importance of the distance between adjacent nanoscale heat sources on the heat dissipation in hotspot systems~\cite{Hoogeboom-Pot4846,PhysRevApplied.11.024042,zeng_disparate_2014,beardo_general_2021,honarvar_directional_2021}.
In 2014, Zeng \textit{et al.}~\cite{zeng_disparate_2014} studied quasiballistic heat conduction for quasi-2D nanoline heat sources, which are periodically deposited on a substrate; based on the frequency-independent phonon Boltzmann transport equation (BTE) under the single-mode relaxation time approximation, they found that the collective behavior caused by closely packed hotspots could counteract the quasiballistic effects in an isolated nanoscale hotspot.
However, the result depends on which temperature signal is used as the fitting data of the diffusion equation.
In 2015, Hoogeboom-Pot \textit{et al.} firstly measured this unexpected phenomenon by advanced dynamic extreme ultraviolet scatterometry~\cite{Hoogeboom-Pot4846}.
To reveal a comprehensive microscopic understanding of this unexpected heat dissipations, in 2021, Honarvar \textit{et al.}~\cite{honarvar_directional_2021} performed the steady-state molecular dynamics simulations on silicon samples featuring close-packed nanoheaters.
They made a qualitative comparison between the molecular dynamics simulations and extreme ultraviolet experiments by controlling for equal ratio between the phonon mean free path and geometry size.
By using atomic-level simulations to accurately access the temperature, phonon scattering and transport properties, they explained that the phonons emitted from the nanoscale heat source may scatter with each other in the in-plane direction and promote the cross-plane heat dissipation when the distance between two nanoscale heat source is smaller than the phonon mean free path.
This heat dissipation phenomenon was also reported by Minnich's research groups by phonon BTE and time-domain thermoreflectance experiments~\cite{PhysRevApplied.10.054068,PhysRevB.97.014307}.
Those results suggest that heat dissipation or cooling in nanoscale hotspot systems including integrated circuits~\cite{warzoha_applications_2021,moore_emerging_2014} might not be as challenging as previously expected.

However, the fundamental physical mechanisms of this novel phenomenon are still not unified.
In addition, it's worth noting that various macroscopic constitutive relationships between the heat flux and temperature are used to fit the experimental data in different research groups~\cite{zeng_disparate_2014,Hoogeboom-Pot4846,PhysRevB.97.014307,beardo_general_2021}.
By artificial fitting, an effective thermal conductivity can be obtained, which varies non-monotonously when the distance between the nanoscale hotspot decreases gradually.
Usually, the heat diffusion equation with a constant effective thermal conductivity is widely used during data post-processing, as did by Hoogeboom-Pot \textit{et al.}~\cite{Hoogeboom-Pot4846} and Zeng \textit{et al.}~\cite{zeng_disparate_2014}, but this model cannot simultaneously fit both amplitude and phase well~\cite{PhysRevApplied.10.054068,beardo_general_2021,PhysRevB.97.014307}.
Under the semi-infinite assumption, Hua and Minnich~\cite{PhysRevB.97.014307} obtained a constitutive relationship between the heat flux and temperature by analytically deriving the phonon BTE under the single-mode relaxation time approximation, which is valid for all phonon transport regimes.
However, this analytical strategy is much challenging for complex geometries and hotspot systems with finite size.
Beardo \textit{et al.} used a macroscopic moment equation with adjustable parameters to fit the experimental data, and both the nonlinear and nonlocal terms of the heat flux are taken into account in their model~\cite{beardo_general_2021}.
They uncovered the existence of two time scales: an interface resistance regime that dominates on short time scales and a quasiballistic phonon transport regime that dominates on longer time scales.
This moment equation is derived from the phonon BTE under the small-perturbation expansion, so that it might be questionable when the systems size is smaller than the phonon mean free path.

Summing up the above, it seems that how to interpret the raw experimental data in the non-diffusive regime with reasonable constitutive relationships is still an open question.
As reported by Zeng \textit{et al.}~\cite{zeng_disparate_2014}, using the temperature signals in different positions for data post-processing might lead to different result.
Hence, it is necessary to obtain the macroscopic physical fields in the whole domain.

Note that there are only a few detection sites in the micro/nano-scale thermal measurement experiments~\cite{Hoogeboom-Pot4846,PhysRevApplied.10.054068,xu_raman-based_2020,honarvar_directional_2021,beardo_observation_2021,ballisticsiliconapl2001,siemens_quasi-ballistic_2010,hu_spectral_2015}, which indicates that it is hard to measure the whole temporal and spatial macroscopic physical fields.
On the other hand, as we all know, heat dissipation in practical thermal engineering span multiple scales of time and space, for example from picoseconds to microseconds or from transistors at the nanoscale to the heat dissipation of a supercomputer~\cite{warzoha_applications_2021}.
Although the molecular dynamics simulations is accurate, it is still too expensive to simulate the dimensions and scales of actual experimental samples or thermal systems.
For example, in Honarvar's work~\cite{honarvar_directional_2021}, the transient extreme ultraviolet experiments is usually at hundreds of nanometers but the steady-state molecular dynamics simulation is below $100$ nanometers.

To the best of our knowledge, the phonon incoherent transport dominates heat conduction in room temperature silicon over tens of nanometers~\cite{MurthyJY05Review,zeng_disparate_2014,PhysRevB.97.014307,PhysRevApplied.10.054068,ma2019b,esee8c149}. Simultaneously considering accuracy and computational efficiency, the phonon BTE simulations are conducted in our work to show the temporal and spatial variations of macroscopic physical fields in the whole 3D finite geometry region.
We mainly focus on how long it takes for the heat to dissipate completely from the heat source.
No artificial fitting or effective thermal conductivity is used to avoid possible controversy caused by data post-processing methods and the raw data calculated by phonon BTE is plotted directly.

The rest of the paper is organized as follows. In Sec.~\ref{sec:bte}, the phonon BTE is introduced. Results and discussions of quasi-2D nanoline (\cref{close_packed2D}) and 3D nanocuboid (\cref{close_packed3D}) hotspot systems are shown in Sec.~\ref{sec:Graynanoline} and~\ref{sec:Graynanocuboid}, respectively. Conclusions are made in Sec.~\ref{sec:conclusion}.

\section{Phonon BTE}
\label{sec:bte}

{\color{red}{In this work, we mainly focused on the heat conduction in conventional 3D semiconductor materials, e.g., monocrystalline silicon and germanium~\cite{PhysRevB.97.014307,terris2009modeling,PhysRevApplied.10.054068,JAP_lACROIXd2014}.
In these materials, the normal process scattering can be ignored and the resistive process scattering dominates the heat conduction~\cite{zeng_disparate_2014,PhysRevB.97.014307,terris2009modeling,LUO2017970,PhysRevApplied.10.054068}. }}
The phonon BTE under the single-mode relaxation time approximation~\cite{GuoZl16DUGKS,LUO2017970,PhysRevB.100.085203,zhang_discrete_2019,MurthyJY05Review,PhysRevB.97.014307,ZHANG20191366} is accurate enough to describe the transient heat conduction in these materials:
\begin{align}
\frac{\partial e}{\partial t }+ v_g \bm{s} \cdot \nabla_{\bm{x}} e  &= \frac{e^{eq}  -e}{\tau  }  ,
\label{eq:BTE}
\end{align}
where $v_g$ is the group velocity and $e=e(\bm{x},\omega,\bm{s},t,p)$ is the phonon distribution function of energy density, which depends on spatial position $\bm{x}$, unit directional vector $\bm{s}$, time $t$, phonon frequency $\omega$ and branch $p$ (Appendix~\ref{sec:dispersionscattering}).
The whole wave vector space is assumed to be isotropic.
$e^{eq} $ and $\tau $ are the equilibrium distribution function and the relaxation time, respectively.
We assume the temperature $T$ slightly deviates from the reference temperature $T_0$, i.e., $|T -T_0| \ll  T_0$, so that the equilibrium distribution function can be linearized as follows:
\begin{align}
e^{eq}_{R}(T) &\approx C \frac{ T-T_0 }{ 4  \pi },  \label{eq:feqR}
\end{align}
where $C=C(\omega,p,T_0)$ is the mode specific heat at $T_0$.
The phonon scattering term satisfies the energy conservation, so that we have
\begin{align}
0 =  \sum_{p} \int \int  \frac{e^{eq}(T_{loc}) -e}{\tau (T_0)}   d\Omega d\omega , \label{eq:Rconsertvation}
\end{align}
where the integral is carried out in the whole solid angle space $d\Omega$ and frequency space $d\omega$.
$T_{loc}$ is the local pseudotemperature, which is introduced to ensure the conservation principles of the scattering term and can be calculated by
\begin{align}
T_{loc} =T_{0}+ \frac{\sum_{p}\int  \frac{\int e d\Omega }{\tau   } d{\omega}   }{ \sum_{p}\int \frac{ C}{\tau  }  d{\omega} } .
\end{align}
The local temperature $T$ and heat flux $\bm{q}$ can be calculated as the moments of distribution function:
\begin{align}
T  &=T_0+ \frac{ \sum_{p} \int \int e d\Omega d\omega } { \sum_{p} \int C d\omega  },   \label{eq:T}  \\
\bm{q} &=  \sum_{p} \int \int \bm{v} e d\Omega d\omega.
\label{eq:heatflux}
\end{align}

\section{Quasi-2D nanoline heat source}
\label{sec:Graynanoline}

\subsection{Problem description}

\begin{figure*}[htb]
\centering
\subfloat[]{\label{close_packed2D}\includegraphics[scale=0.5,clip=true]{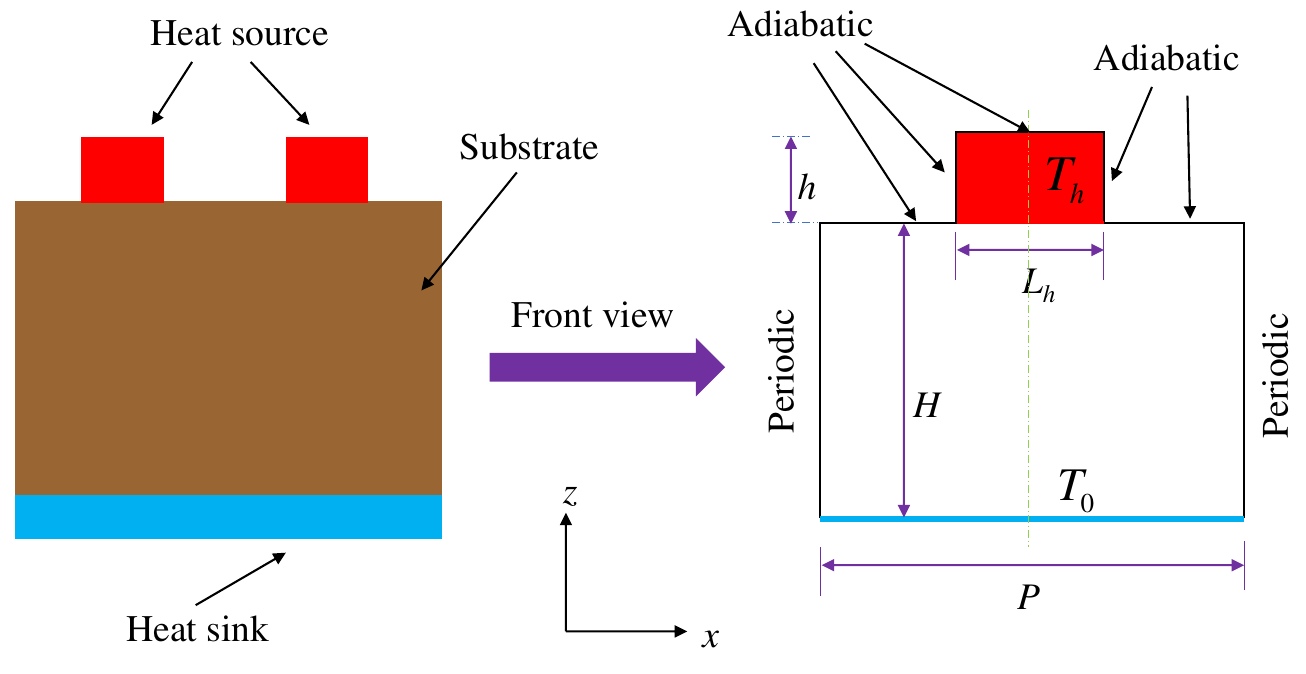} } \\
\centering
\subfloat[]{\includegraphics[scale=0.35,clip=true]{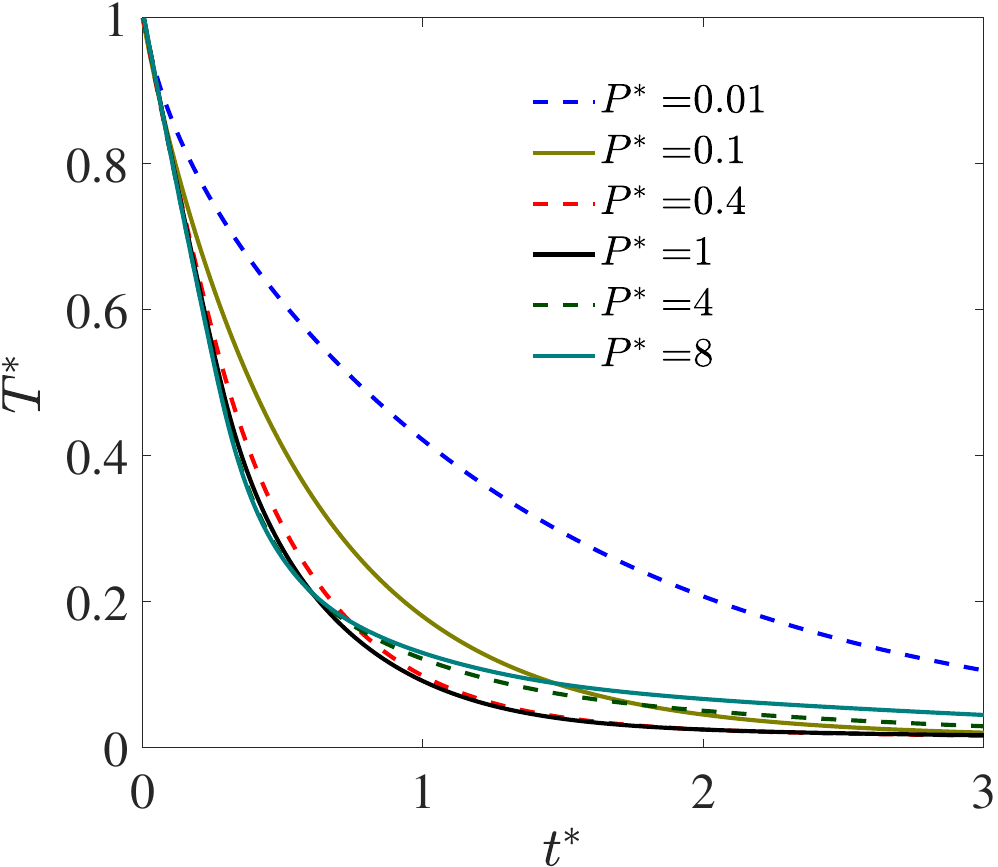}}~~~
\subfloat[]{\includegraphics[scale=0.35,clip=true]{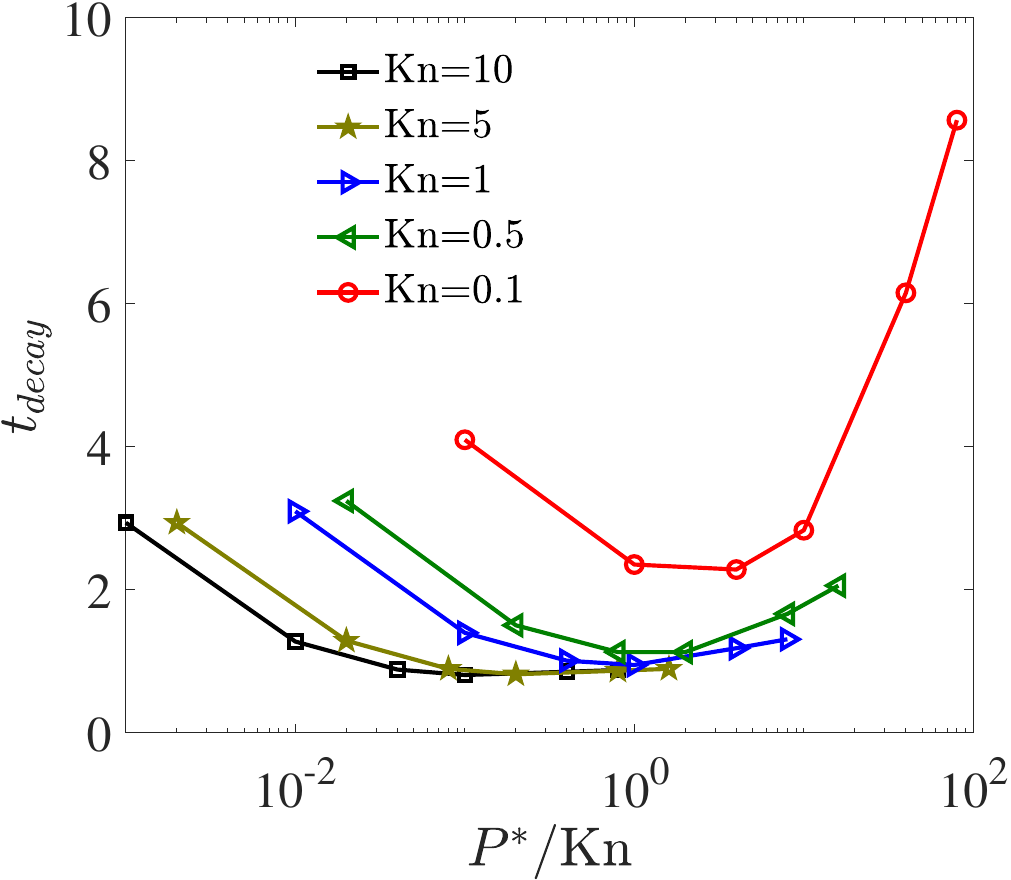}} \\
\caption{(a) Schematic of the transient heat dissipation in quasi-2D nanoline heat source with periodic array arrangement. (b) Heat dissipation process of the average temperature \eqref{eq:dimensionlessT} based on gray model, where $\text{Kn}= 1.0$. (c) The time decay $t_{decay}$ with various $P^*$ and $\text{Kn}$.  }
\label{nanoline2DR}
\end{figure*}

The heat dissipations in quasi-2D nanoline hotspot systems are investigated numerically. As shown in~\cref{close_packed2D},
a heat source is added on the top of a rectangle substrate and its sizes in the $x$ and $z$ directions are $L_h$ and $h$, respectively.
The sizes of the substrate in the $x$ and $z$ directions are $P$ and $H$, respectively.
The bottom of the substrate is the heat sink with environment temperature $T_0$ and the isothermal boundary condition is used (Eq.~\eqref{eq:isothermal}).
The left and right boundaries of the substrate are periodic and the others are diffusely reflecting adiabatic boundaries (Eq.~\eqref{eq:diffusely}).
We fix $h/H=1/8$, $L_h /P =1/4$, and the whole domain is a homogeneous material in order to eliminate the thermal interface resistance between two dissimilar materials~\cite{RevModPhysTBR,RevModPhys.94.025002}.

At the initial moment $t=0$, the temperature of the heat source and the other areas are $T_h$ and $T_0$, respectively, where $T_h> T_0$.
When $t>0$, the heat dissipates from the heat source to the heat sink.
The temporal evolutions of the average temperature $\overline{T}$ are studied based on phonon BTE:
\begin{align}
T^*= \frac{ \overline{T} - T_0 }{T_h  -T_0},
\label{eq:dimensionlessT}
\end{align}
where $\overline{T}$ is the average temperature over the whole heat source.
We mainly focus on how long it takes for heat to dissipate completely from the heat source. {\color{red}{Specifically, we study the factors influence the time decay $t_{decay}$, which is defined as the time cost when $T^*$ decreases from $1.0$ to $0.1$.}}

\begin{figure*}[htb]
\centering
\subfloat[]{\includegraphics[scale=0.5,viewport=0 0 400 250,clip=true]{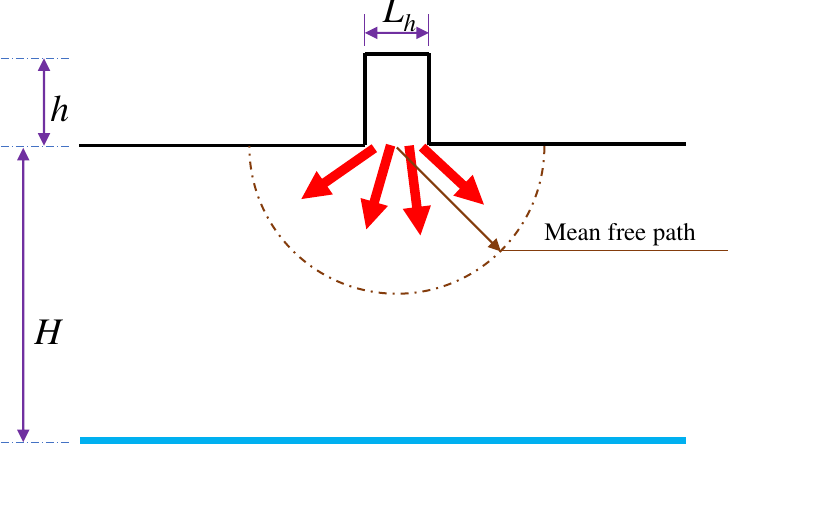} }~~
\subfloat[]{\includegraphics[scale=0.5,viewport=420 0 880 250,clip=true]{physics_packed2D.pdf} }~~
\caption{Schematic of phonon transport and scattering in (a) single hotspot and (b) close-packed hotspot systems. }
\label{physics_packed2D}
\end{figure*}
Based on dimensional analysis~\cite{barenblatt1987dimensional}, the transient heat dissipations in the quasi-2D nanoline hotspot systems are totally determined by these length scales, including the phonon mean free path $\lambda = v_g \tau $, the spatial period $P$, height $H$ and the size of hotspot $L_h$.
Equation~\eqref{eq:BTE} can be written in the dimensionless form:
\begin{align}\label{eq:dimensionlessBTE}
\frac{\partial e}{\partial t }+ \bm{s} \cdot \nabla_{\bm{x}} e  &= \frac{e^{eq}  -e}{\text{Kn}  }  ,
\end{align}
where the distribution function is normalized by $e_{\text{ref}}  ={ C  \Delta T  }/{ (4 \pi)}$ with $\Delta T= T_h -T_0$ being the temperature difference in the domain, the spatial coordinates normalized by $H$, and time normalized by $t_{ref}=H/v_g $.
The dimensionless Knudsen number is
\begin{equation} \label{eq:dimensionlessparameters}
\text{Kn}^{-1} = \frac{H}{ \lambda  }= \frac{H}{ v_g \tau  }
\end{equation}
In order to better pinpoint the relationships among various influencing factors, two dimensionless parameters are introduced and defined as
\begin{align}
P^*=\frac{P}{H}, \quad  t^*= \frac{v_g t}{H}.
\end{align}

\subsection{Effects of geometric sizes and phonon scattering}

The phonon gray model~\cite{MurthyJY05Review,zeng_disparate_2014} and the linear phonon dispersion are used.
In this simulation, the height $H$ is fixed.
The detailed numerical solutions of BTE are shown in Appendix~\ref{sec:solver} as well as the independence test conducted in Appendix~\ref{sec:validation}.

The thermal effects of the spatial period $P$ are investigated.
As shown in~\cref{nanoline2DR}(b) with $\text{Kn}=1.0$, the heat dissipation efficiency is not monotonic when $P^*$ decreases from $8$ to $0.01$.
The time decay $t_{decay}$ is also plotted in~\cref{nanoline2DR}(c).
When $P^* =1.0$ or $0.4$, the heat dissipation speed is the fastest.
Note that both $v_g$ and $H$ are fixed when the spatial period $P$ changes, so that the dimensionless time $t^*$ is equivalent to the actual physical time $t$.

Next, a number of simulations are carried out with various $\text{Kn}$.
It can be found that the non-monotonic heat dissipation phenomenon still exists with different Knudsen numbers.
The present results clearly contradict previous intuitive understanding of micro/nano scale heat transfer, namely, the more densely packed and smaller the electronics, the more difficult it is to dissipate heat~\cite{warzoha_applications_2021,moore_emerging_2014}.

\subsection{Physical mechanisms}
\label{sec:mechsnisms}

{\color{red}{Motivated by previous studies of quasiballistic phonon transport~\cite{chen1996,zeng_disparate_2014,Hoogeboom-Pot4846,honarvar_directional_2021,PhysRevApplied.10.054068,PhysRevB.97.014307}, the fundamental physical mechanisms of above unexpected thermal transport phenomena in different phonon transport regimes are discussed qualitatively.}}
From~\cref{close_packed2D} or~\cref{physics_packed2D}, it can be found that there are two main thermal transport processes when heat is transferred from the heat source to the heat sink~\cite{zeng_disparate_2014,Hoogeboom-Pot4846}: phonons escape from the heat source to the substrate and phonons transport from the substrate to the heat sink.
Based on dimensionless analysis~\cite{barenblatt1987dimensional}, for the first process, the size of the heat source is the key factor, especially $L_h/h$.
For the second process, namely, phonons with high energy are absorbed by the heat sink, the distance $P$ between nanoscale heat source and height $H$ determine the heat dissipation efficiency.
In addition, the phonon group velocity and relaxation time influence both two transient heat dissipation processes.

Diffusive.—When the spatial period is much larger than the phonon mean free path, $P \gg \lambda $ and $L_h \gg  \lambda   $, the phonon scattering is very sufficient inside both the heat source and substrate areas and phonons undergo a diffusive process. Hence, the Fourier's law is valid and the temperature decreases exponentially.

Quasiballistic.—When the spatial period decreases and becomes comparable to the phonon mean free path, the thermal dissipation mechanisms become much complicated.
For the first process, $L_h/h$ decreases so that it becomes difficult for phonons to escape from the heat source areas.
{\color{red}{For the second process, if there is only a single nanoscale heat source, as shown in~\cref{physics_packed2D}(a), when phonons escape from the heat source, there is rare phonon-phonon scattering within the spatial range of a phonon mean free path.}}
The insufficient phonon scattering blocks the efficient energy exchange among phonons and a large thermal resistance appears near the outlet position of the heat source~\cite{chen1996,chen_non-fourier_2021}.

\begin{figure}[htb]
\centering
\includegraphics[scale=0.35,clip=true]{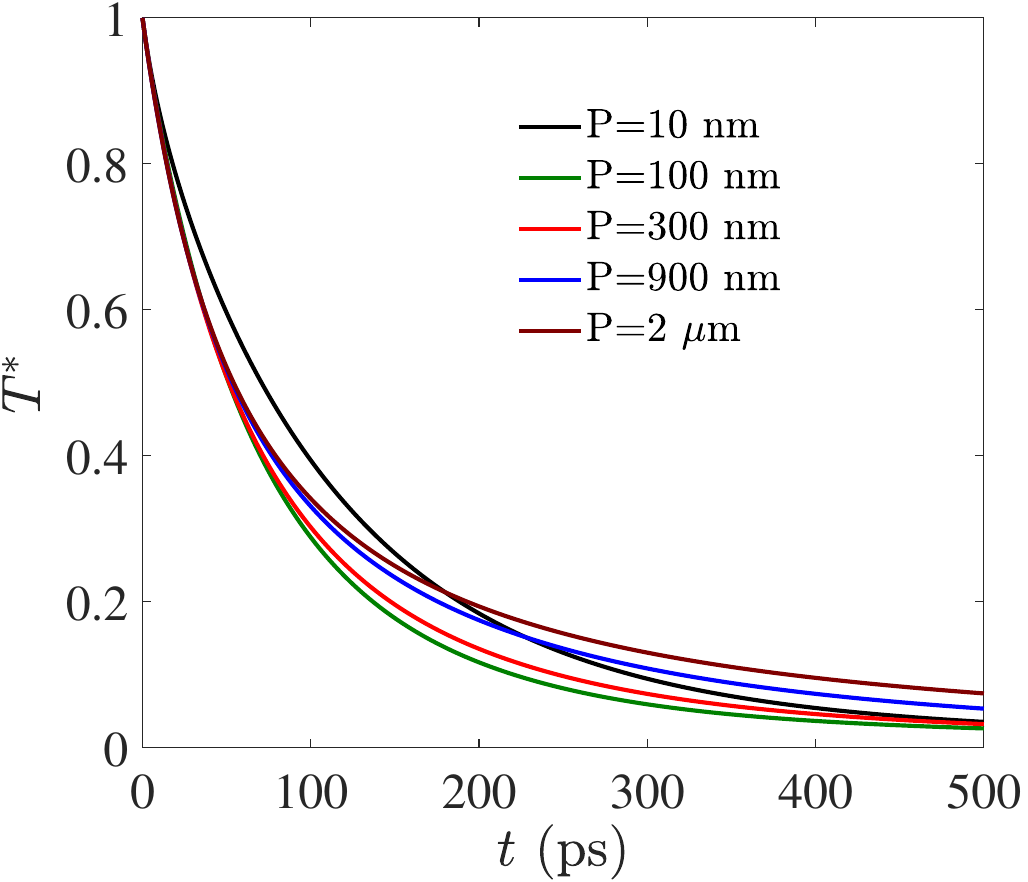} \\
\includegraphics[scale=0.3,clip=true]{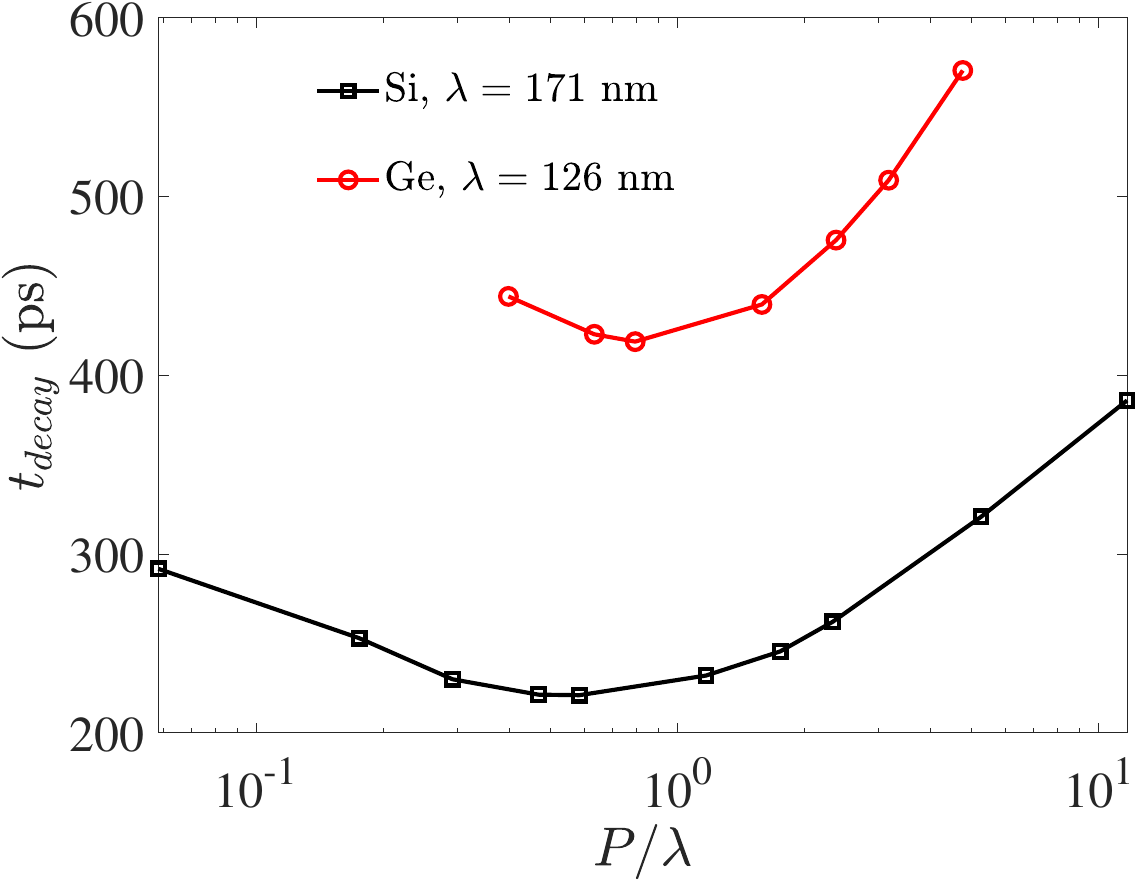}
\caption{
(a) Heat dissipation process of the average temperature \eqref{eq:dimensionlessT} in silicon materials with quasi-2D nanoline geometry (\cref{close_packed2D}) based on frequency-dependent BTE, where $ H= 300$ nm, $T_0 =300$ K. (b) The time decay $t_{decay}$ with various $P$ in silicon and germanium materials.
  }
\label{nanolinesilicon}
\end{figure}
\begin{figure*}[htb]
\centering
\subfloat[]{ \label{close_packed3D} \includegraphics[scale=0.5,clip=true]{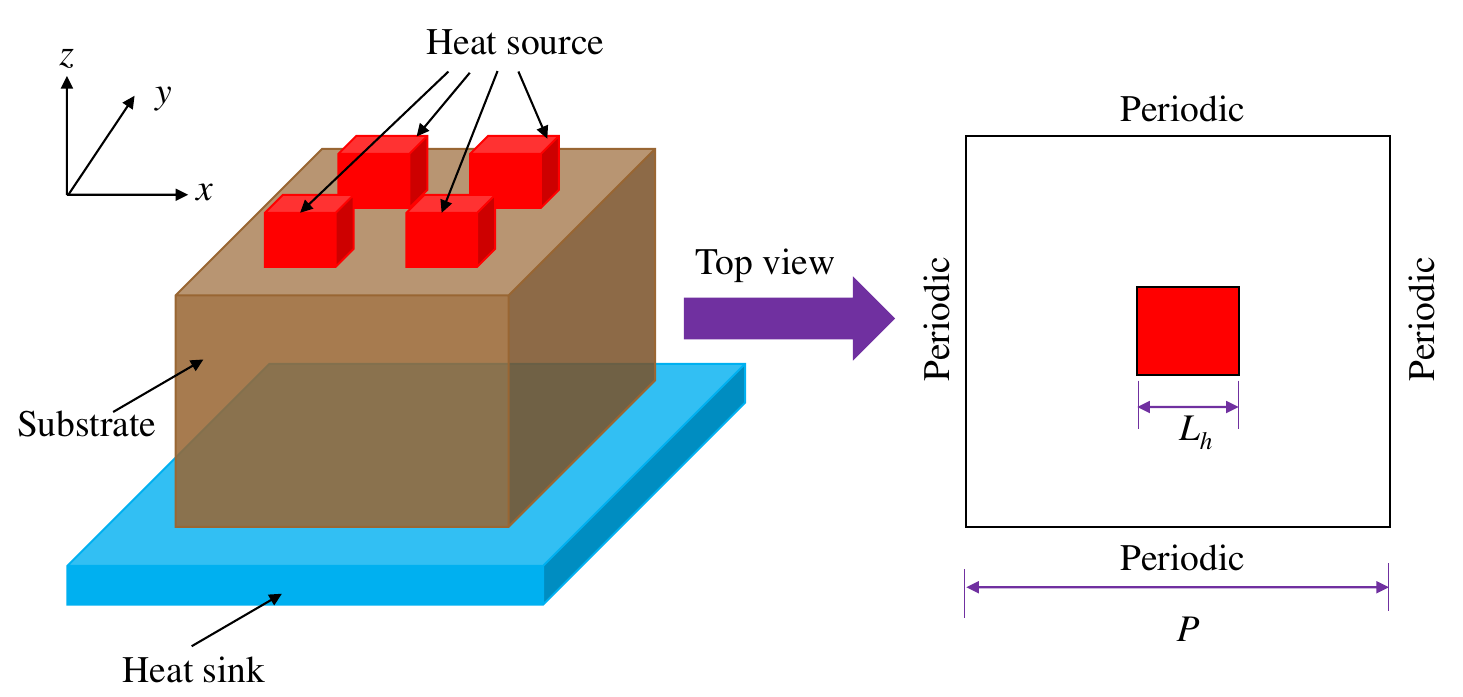} }  \\
\subfloat[]{\includegraphics[scale=0.35,clip=true]{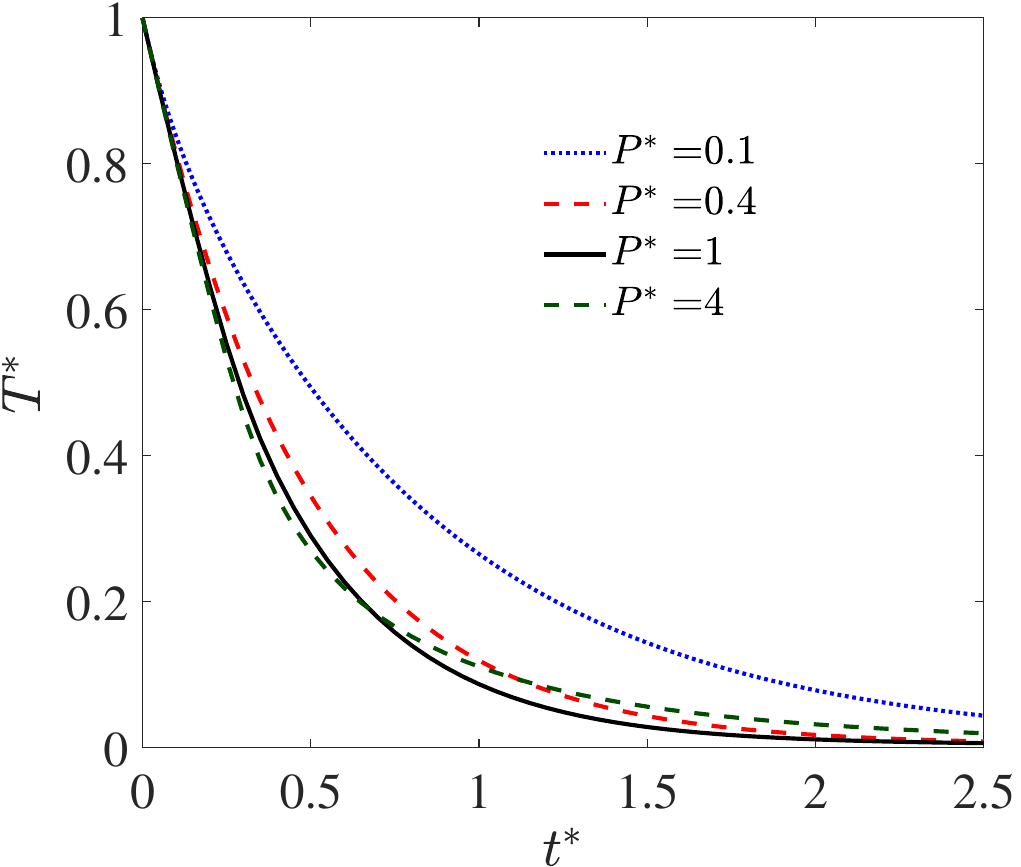}}~~~
\subfloat[]{\includegraphics[scale=0.35,clip=true]{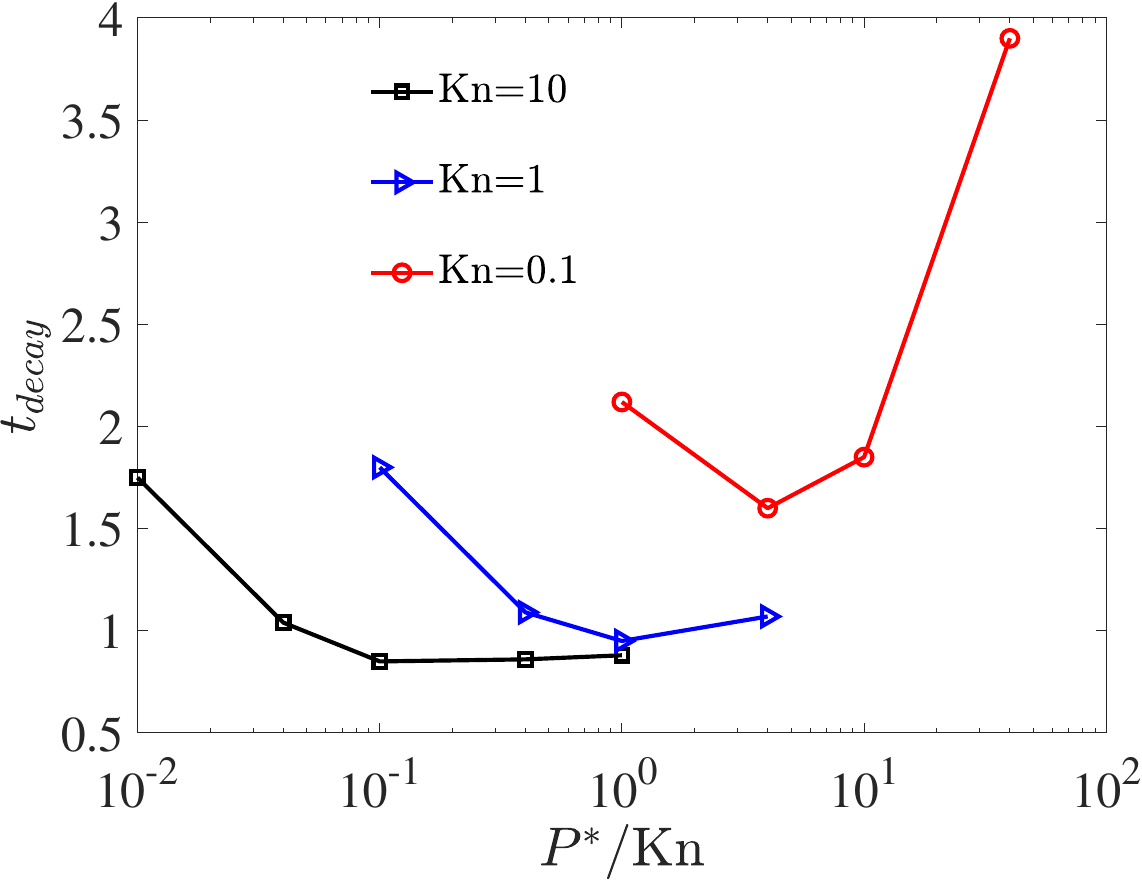}}  \\
\caption{
(a) Schematic of the transient heat dissipation in 3D nanocuboid heat source with periodic array arrangement. (b) Heat dissipation process of the average temperature \eqref{eq:dimensionlessT} based on gray model, where $\text{Kn}=1$. (c) The time decay $t_{decay}$ with various $P^*$ and $\text{Kn}$.
 }
\label{nanocuboid3D}
\end{figure*}
\begin{figure*}[htb]
\centering
\subfloat[]{ \includegraphics[scale=0.5,clip=true]{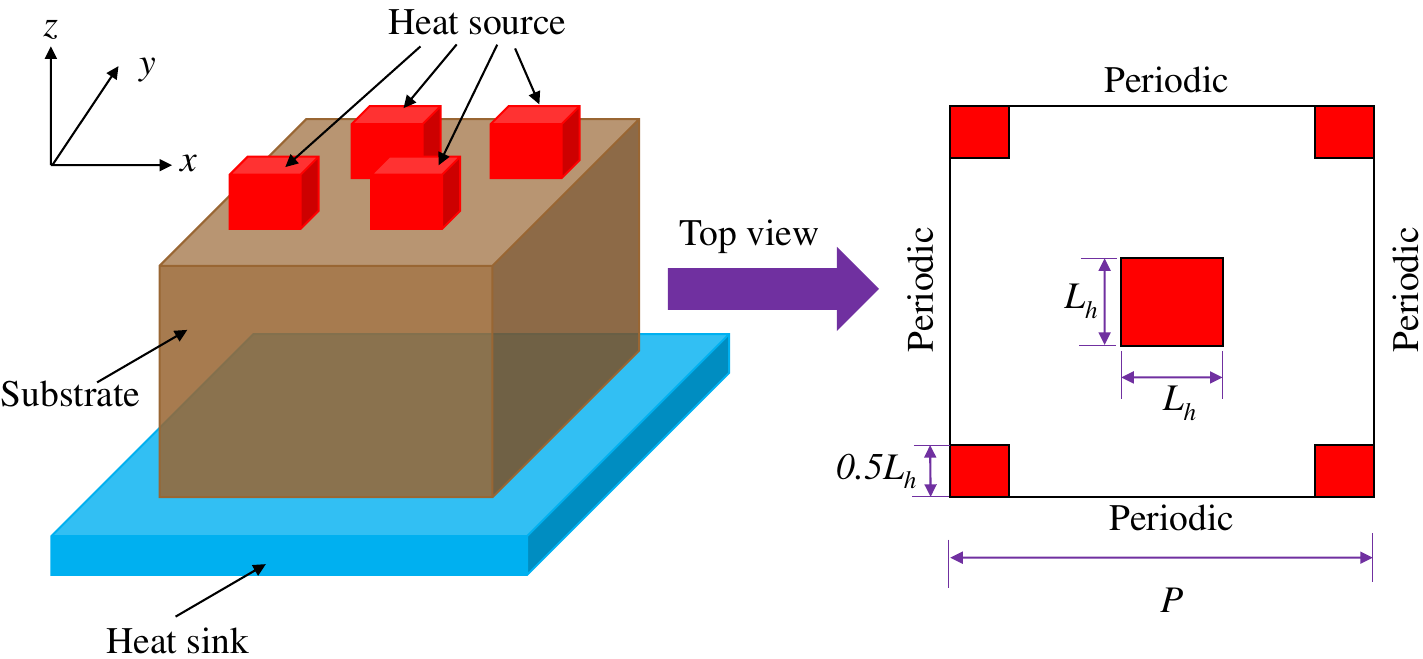} }  \\
\subfloat[]{\includegraphics[scale=0.35,clip=true]{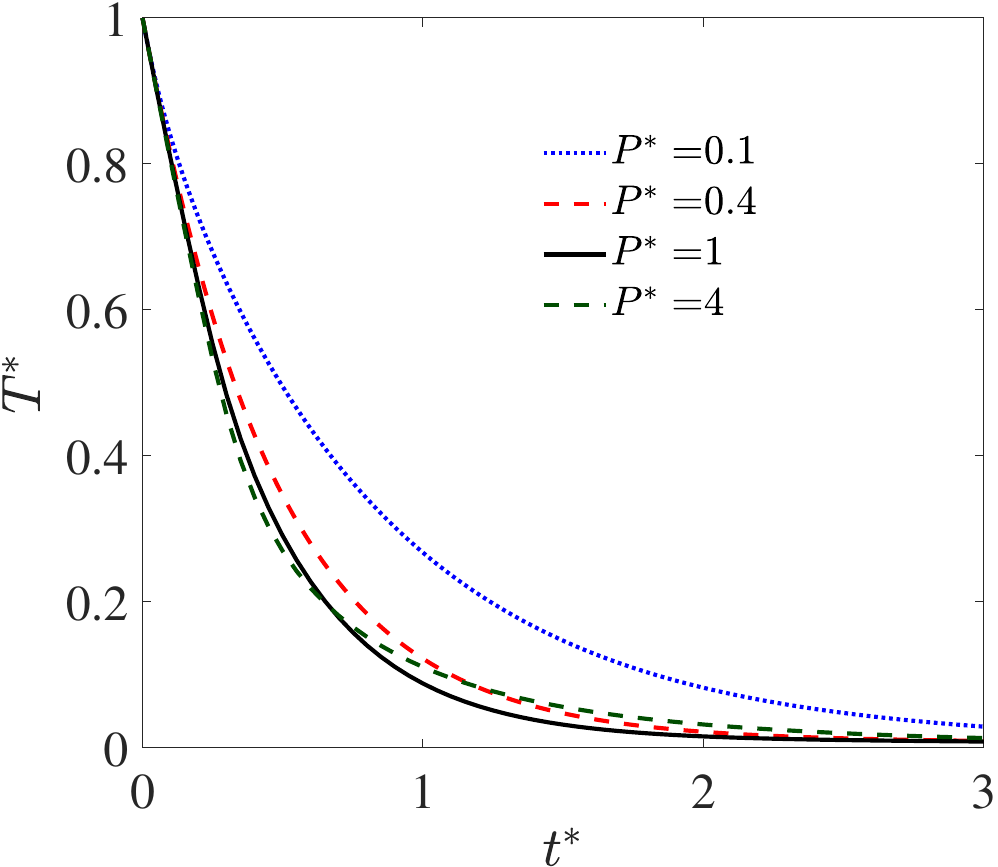}}~~~
\subfloat[]{\includegraphics[scale=0.35,clip=true]{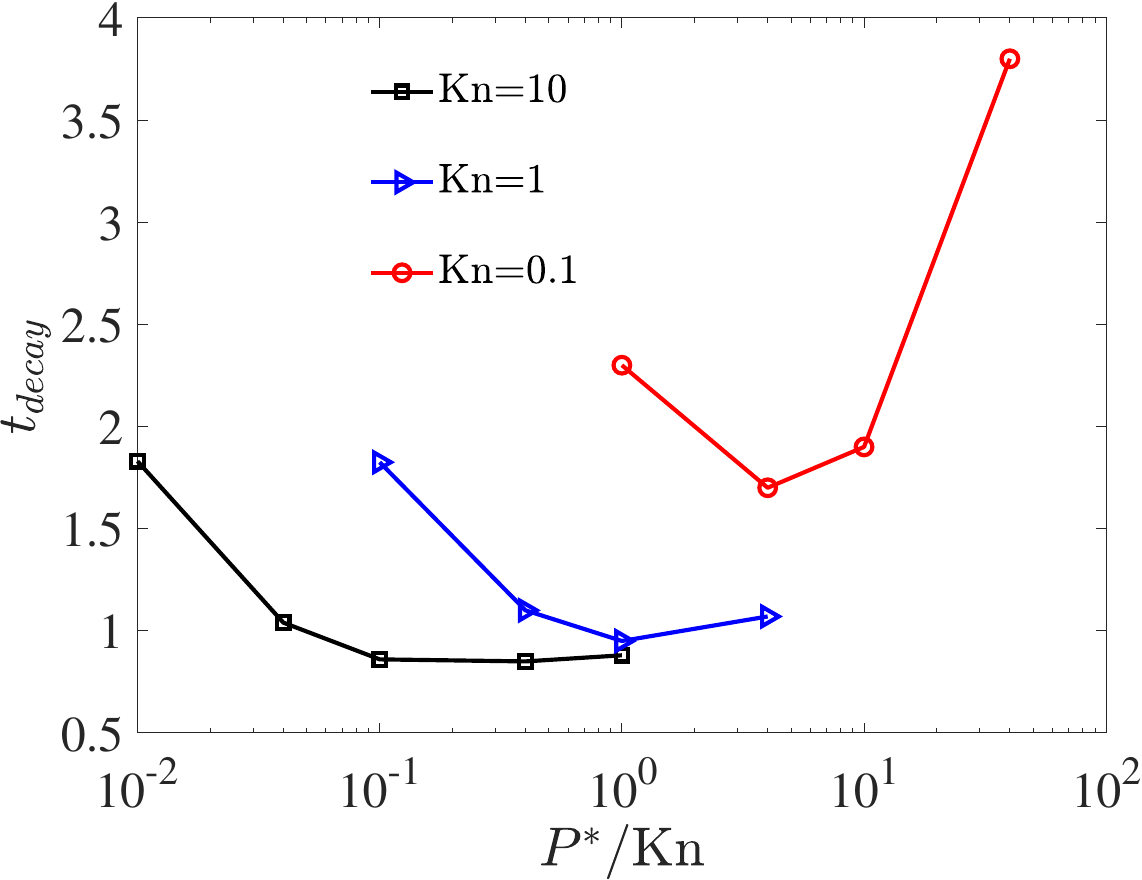}}  \\
\caption{
(a) Schematic of the transient heat dissipation in 3D nanocuboid heat source with periodic staggered arrangement. (b) Heat dissipation process of the average temperature \eqref{eq:dimensionlessT} based on gray model, where $\text{Kn}=1$. (c) The time decay $t_{decay}$ with various $P^*$ and $\text{Kn}$.
 }
\label{nanostaggered3D}
\end{figure*}
When a number of heat sources are periodically deposited on a substrate, it should be noted that the distance between two nanoscale heat source decreases if $P$ decreases.
The phonons escaped from one nanoscale heat source may scatter with others escaped from the adjacent heat source.
{\color{red}{In other words, when the distance between two nanoscale heat source decreases and $P \approx \lambda$, compared to that with a single nanoscale hotspot, the phonon-phonon scattering is instead boosted within the spatial range of a phonon mean free path, as shown in~\cref{physics_packed2D}(b).}}
The heat flux in the $x$ direction is canceled out by phonons coming from opposite directions.
And the heat conduction in the $z$ direction is increased, which is totally different from that of a single nanoscale heat source~\cite{Hoogeboom-Pot4846,honarvar_directional_2021}.

$Ballistic$.—When the spatial period is much smaller than the phonon mean free path, $P \ll \lambda   $ and $L_h \ll \lambda $, the ballistic phonon transport dominates heat conduction inside both the heat source and substrate areas.
Although the smaller distance between two nanoscale heat source could promote scattering, the ratio $L_h/h$ decreases significantly so that the phonon transport is blocked by the diffusely reflecting boundaries and it is much difficult for most of heat/phonons to escape from the heat source to the substrate areas.
In other words, the first process totally dominates phonon transport and limits the heat dissipation.

{\color{red}{Combined with our numerical results in~\cref{nanoline2DR} and the theoretical analysis of quasiballistic phonon transport~\cite{zeng_disparate_2014,Hoogeboom-Pot4846,honarvar_directional_2021}, it is concluded that in quasi-2D hotspot systems, the heat dissipation efficiency reaches the highest value when $P^* /\text{Kn} \approx 1$.
It is a competition result of two phonon transport processes: quasiballistic transport when phonons escape from the nanoscale heat source and the scattering among phonons originating from adjacent nanoscale heat source.}}

\subsection{Silicon and germanium materials}

The quasi-2D nanoline hotspot systems (\cref{close_packed2D}) with room temperature monocrystalline silicon and germanium material are studied based on frequency-dependent phonon BTE.
{\color{blue}{The input parameters of BTE including nonlinear phonon dispersion and frequency-dependent scattering processes are given in Appendix~\ref{sec:dispersionscattering}, which have been validated by experiments in previous studies~\cite{Glassbrenner64conductivity,terris2009modeling,JAP_lACROIXd2014}.}}
The average phonon mean free path $ \lambda = \left( \sum_{p} \int C v_g^2 \tau  d\omega \right) /\left( \sum_{p} \int C v_g d\omega \right)  $ of room temperature silicon is about $171$ nm.
The thermal effects of the spatial period $P$ on the heat dissipation are investigated, and the height is fixed at $H=300$ nm~\cite{honarvar_directional_2021}.
From~\cref{nanolinesilicon}(a)(b), it can be found that in silicon the heat dissipation efficiency is low when $P=2~\mu$m or $10$ nm, and the efficiency is the fastest when $ P \approx 100$ nm.
{\color{blue}{Similar non-monotonic heat dissipation phenomena are also observed in germanium materials with the average mean free path $\lambda=126$ nm, see~\cref{nanolinesilicon}(b).}}
{\color{red}{These results are consistent with our analysis in Sec.~\ref{sec:mechsnisms}, namely, the heat dissipation efficiency reaches the highest value when the spatial period $P$ is approximately the phonon mean free path $\lambda$, i.e., $P^*  \approx \text{Kn}$.}}

\section{3D nanocuboid heat source}
\label{sec:Graynanocuboid}

The 3D close-packed nanocuboid heat source is simulated in this section.
As shown in~\cref{close_packed3D}, a number of nanocuboid heat source are arranged periodically on the top of the substrate.
The bottom of the 3D geometry is the heat sink with fixed temperature $T_0$ and the isothermal boundary condition is used (Eq.~\eqref{eq:isothermal}).
Its front and left views are both the same as the front view plotted in~\cref{close_packed2D}.
The boundaries of the heat source and the top surface of the substrate are diffusely reflecting adiabatic boundaries (Eq.~\eqref{eq:diffusely}).
From the top view, there are two concentric squares with side length $P$ and $L_h$, and the boundaries of the substrate are all periodic.
The length of the substrate and nanocuboid in the $z$ direction is $H$ and $h=H/8$, respectively.
The basic settings are similar to those in quasi-2D hotspot systems (\cref{close_packed2D}).
At the initial moment $t=0$, the temperature of the heat source is $T_h $ and the temperature of the other surfaces is $T_0$.
When $t>0$, the heat dissipates from the heat source to the heat sink.

The detailed numerical solutions are shown in Appendix~\ref{sec:solver}.
Due to the large computational amount, less numerical cases are simulated compared to those in quasi-2D hotspot systems, and the frequency-independent BTE is solved.
The thermal effects of phonon scattering and spatial period $P$ are investigated.
From~\cref{nanocuboid3D}, it can be found that the heat dissipation phenomena are similar to those in~\cref{nanoline2DR}.
Namely, there is non-monotonic heat dissipation phenomenon when the distance between two adjacent nanoscale hotspot decreases gradually.
{\color{red}{The fastest heat dissipation speed appears when $P^* \approx  \text{Kn} $. }}

%The reason is discussed qualitatively as below.
%Figure~\ref{physics_packed2D} and Sec.~\ref{sec:mechsnisms} have explained that the non-monotonic heat dissipation phenomenon results from the scattering among phonons originating from adjacent nanoscale heat source, which has also been reported in previous studies~\cite{zeng_disparate_2014,Hoogeboom-Pot4846,honarvar_directional_2021,PhysRevApplied.10.054068,PhysRevB.97.014307}.
%But there are something different in quasi-2D and 3D hotspot systems.
%In this study, the height $H$ is fixed so that the top view of the quasi-2D and 3D geometries are shown in~\cref{top_view}.
%In quasi-2D structure, there is only single heat dissipation direction, which are affected significantly by the spatial period $P$ and phonon mean free path $\lambda$, as shown in~\cref{top_view}(a).
%Then the possibility of the scattering among phonons originating from adjacent nanoscale heat source is related to $P^*/ \text{Kn}$.
%However, in 3D structure, there are two heat dissipation directions.
%The phonons could scatter with other phonons originating from left (right) or top (bottom) nanoscale heat source, as shown in~\cref{top_view}(b).
%Then the possibility of the scattering among phonons originating from adjacent nanoscale heat source is different from that in quasi-2D geometry, and it is related to $(P^*)^2/ \text{Kn}$, rather than $P^*/ \text{Kn}$}}

{\color{blue}{In addition, changing the spatial distributions of nanocuboid heat sources, from periodic array to staggered arrangement, as shown in~\cref{nanostaggered3D}, numerical results show that this non-monotonic heat dissipation phenomenon still exists.}}
Thus, it can be concluded that the non-monotonic heat dissipation phenomena are general in both close-packed quasi-2D and 3D hotspot systems.

\section{Conclusion}
\label{sec:conclusion}

In summary, the heat dissipation in close-packed quasi-2D nanoline and 3D nanocuboid hotspot systems has been studied based on phonon BTE.
Against the previous intuitive understanding of micro/nano scale heat conduction, the present results have revealed that the heat dissipation efficiency is not monotonic with the distance between heat sources.
{\color{red}{The highest heat dissipation efficiency is reached when $P^* /\text{Kn} \approx 1$.}}
It is a competition result of two processes: quasiballistic phonon transport when phonons escape from the nanoscale heat source and the scattering among phonons originating from adjacent nanoscale heat source.
%In the future, the heat dissipation in practical electronic devices or electric vehicles with thermal interface resistance will be studied~\cite{RevModPhys.94.025002,yue_advances_2021,RevModPhysTBR,warzoha_applications_2021,moore_emerging_2014}.

\section*{Acknowledgments}

This work is supported by the National Natural Science Foundation of China (12147122) and the China Postdoctoral Science Foundation (2021M701565).
The authors acknowledge Chengyun Hua and Albert Beardo Ricol for useful communications on quasiballistic phonon transport.
The computational resource is supported by the Center for Computational Science and Engineering of Southern University of Science and Technology.

\appendix

{\color{blue}{
\section{Nonlinear phonon dispersion and scattering}
\label{sec:dispersionscattering}

The thermal contribution of optical phonons are small in room temperature silicon/germanium so that we only consider the longitudinal and transverse acoustic (LA and TA) phonons.
The Pop's formulas~\cite{pop2004analytic} are used to express the isotropic dispersion relations of the acoustic phonon branches,
\begin{equation}
\omega=c_{1}k+c_{2}k^2,
\label{eq:curves}
\end{equation}
where the wave vector $k \in[0,k_{\text{max}}]$, $k_{\text{max}}=2\pi /A $ is the maximum wave vector in the first Brillouin zone, $A$ is the lattice constant.
For silicon, $A =0.543$ nm, and for germanium, $A=0.565$ nm.
The value of the group velocity is $v_g = c_{1} +2 c_{2} k$.
The specific values of these coefficients in Eq.~\eqref{eq:curves} are shown in Table~\ref{dispersionSiGe}.

The Matthiessen's rule is used to calculate the effective relaxation time, $\tau ^{-1}=\tau_{{\text{impurity}}}^{-1}+\tau_{{\text{U}}}^{-1}+\tau_{{\text{N}}}^{-1}=\tau_{{\text{impurity}}}^{-1}+\tau_{{\text{NU}}}^{-1}$, where $\tau_{{\text{impurity}}}^{-1}  =  A_{i}\omega^{4}$.
For LA branch, $\tau_{{\text{NU}}}^{-1}=B_{L}\omega^{2}T^{3}$;
For TA branch, when $0 \leq k < k_{max}/2$, $\tau_{{\text{NU}}}^{-1}=B_T\omega T^4$ and when $k_{max}/2 \leq k \leq k_{max}$, $\tau_{{\text{NU}}}^{-1}=B_U\omega^{2}/{\sinh(\hbar\omega/k_{B}T)}$.
The specific values of these coefficients of relaxation time are shown in Table~\ref{relaxationparameters}.

\renewcommand\arraystretch{1.5}
\begin{table}
\caption{Quadratic phonon dispersion coefficients for silicon (Si) and germanium (Ge)~\cite{pop2004analytic,terris2009modeling,JAP_lACROIXd2014}. }
\centering
\begin{tabular}{|*{4}{c|}}
\hline
 &  $c_{1}$ ($10^5$ cm/s)  & $c_{2}$ ($10^{-3}$ $\text{cm}^{2}$/s)  \\
 \hline
Si, LA  &  9.01 & -2.0      \\
 \hline
Si, TA  &  5.23 & -2.26      \\
 \hline
Ge, LA  &  5.63 & -1.5      \\
 \hline
Ge, TA  &  2.60 & -1.13      \\
\hline
\end{tabular}
\label{dispersionSiGe}
\end{table}
\begin{table}
\caption{Relaxation time coefficients for silicon and germanium~\cite{JAP_lACROIXd2014,terris2009modeling}. }
\centering
\begin{tabular}{|*{3}{c|}}
\hline
&  Silicon  & Germanium \\
\hline
$A_{i}$ (${\text{s}^{\text{3}}}$) &  $ 1.498\times10^{-45} $ &  $2.40 \times  10^{-44} $       \\
\hline
$B_{L}$ (${\text{K}^{\text{-3}}}$) & $ 1.180\times 10^{-24} $ &  $ 2.30 \times 10^{-24}   $     \\
\hline
$B_T$ (${\text{K}^{\text{-3}}}$)  & $ 8.708\times 10^{-13}  $ &  $ 3.0 \times  10^{-12}   $  \\
\hline
$B_{U}$ (${\text{s}}$) &  $ 2.890 \times 10^{-18} $  &  $1.50  \times  10^{-18}$   \\
\hline
\end{tabular}
\label{relaxationparameters}
\end{table}

These input parameters of the phonon BTE have been validated by experiments~\cite{Glassbrenner64conductivity} in previous studies~\cite{terris2009modeling,JAP_lACROIXd2014}.
}}

\section{Numerical method for BTE}
\label{sec:solver}
\begin{figure*}[htb]
\centering
\subfloat[$\text{Kn}=10$]{\includegraphics[scale=0.3,clip=true]{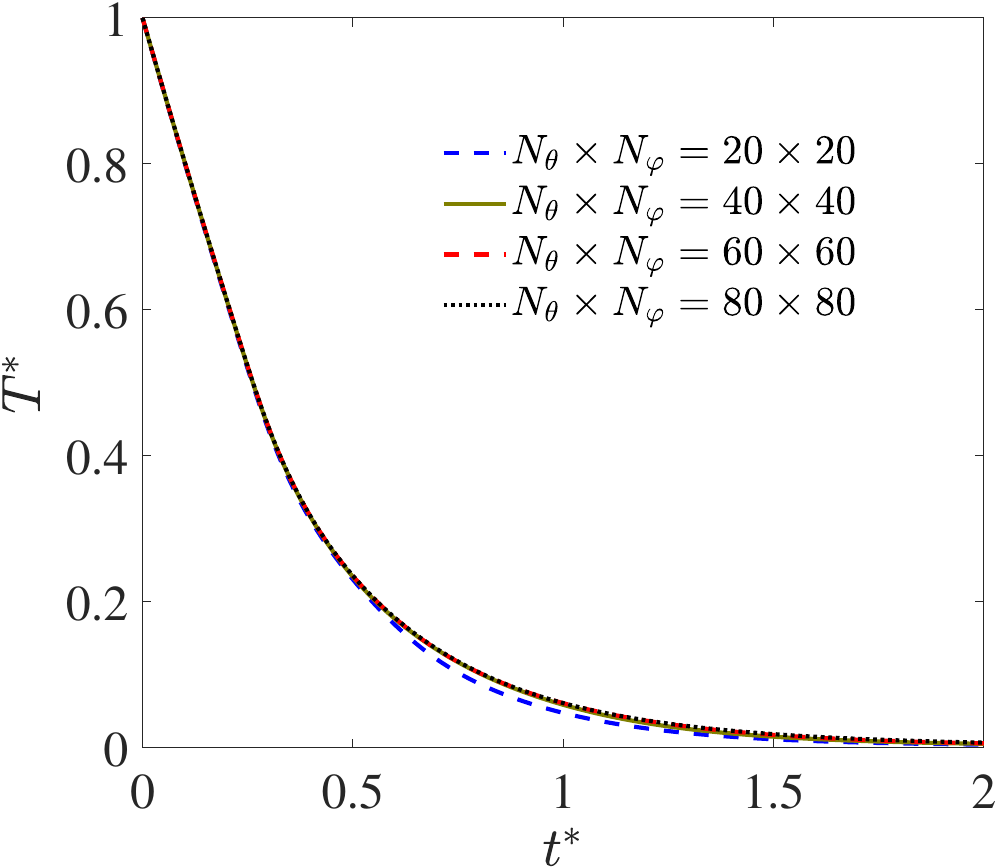}}~
\subfloat[$\text{Kn}=10$]{\includegraphics[scale=0.3,clip=true]{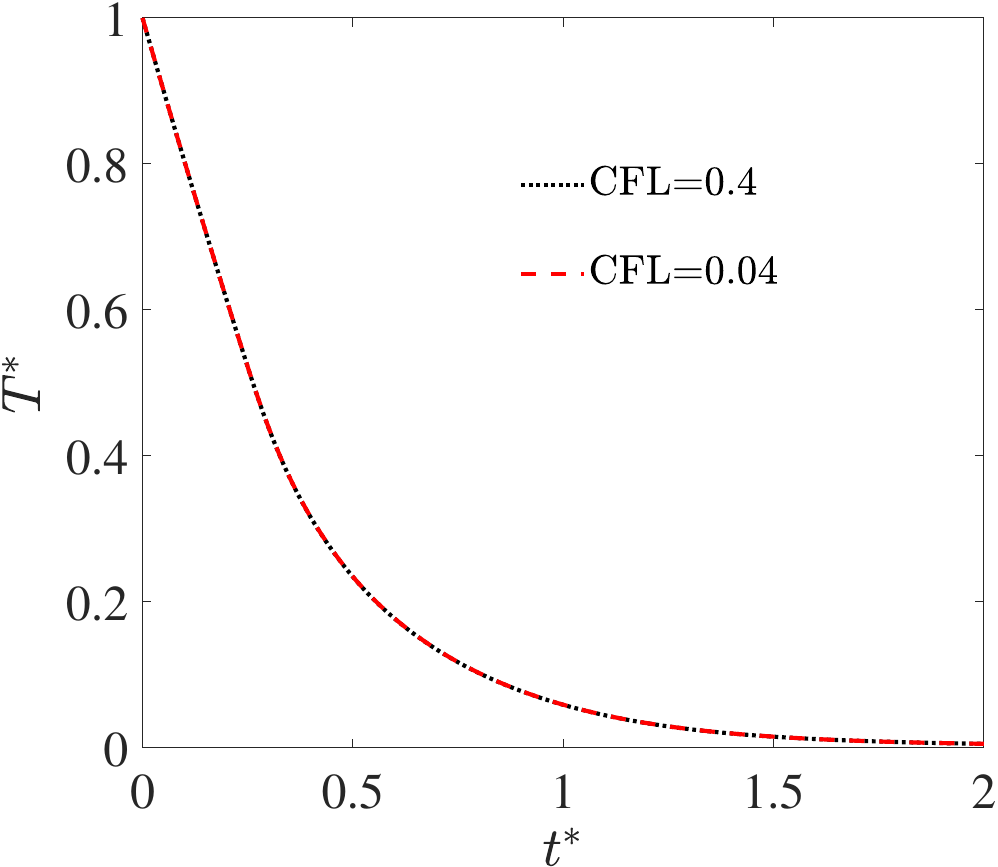}} ~
\subfloat[$\text{Kn}=0.1$]{\includegraphics[scale=0.3,clip=true]{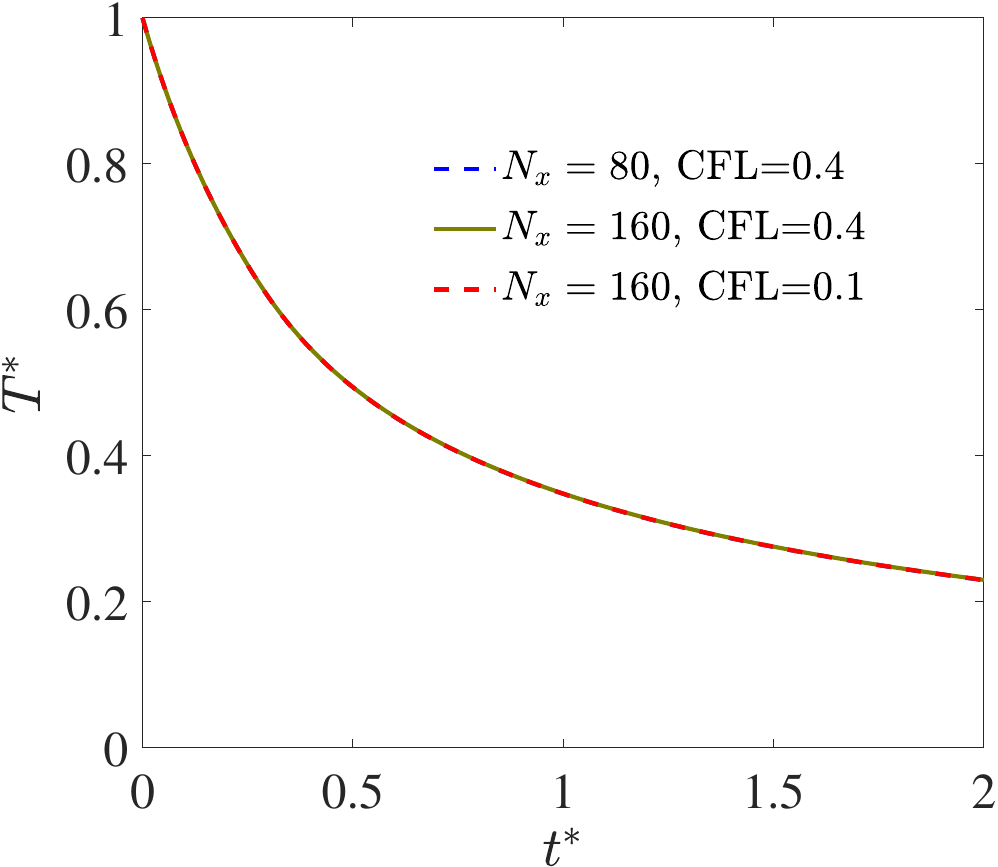}}
\caption{Independence tests of the discrete parameters. Heat dissipation process of the average temperature \eqref{eq:dimensionlessT} in quasi-2D hotspot system, when $P^* =1 $. (a) Different numbers of the discretized solid angles $N_{\theta} \times N_{\varphi}$, fixed discretized cells in the $x$ direction $N_x =80$ and fixed CFL number $\text{CFL} =0.4$.  (b) Different CFL number, fixed $N_{\theta} \times N_{\varphi}=40 \times 40$ and fixed $N_x =80$. (c) Different CFL number and $N_x$, fixed $N_{\theta} \times N_{\varphi}=40 \times 40$. }
\label{Independence_tests}
\end{figure*}

The discrete unified gas kinetic scheme~\cite{guo_progress_DUGKS} is used to solve the phonon BTE numerically.
Detailed introductions and numerical validations of this scheme are given in Refs.~\cite{GuoZl16DUGKS,LUO2017970,zhang_discrete_2019}.
For quasi-2D nanoline hotspot systems, the spatial space is discretized with $90$ uniform cells in the $z$ direction and $40\sim200$ uniform cells in the $x$ direction.
In silicon or germanium materials, the spatial space is discretized with $N_z= 90$ uniform cells in the $z$ direction and $N_x= 40\sim120$ uniform cells in the $x$ direction.
For the 3D nanocuboid hotspot systems, similarly, the spatial space is discretized with $N_z= 90$ uniform cells in the $z$ direction and $N_x=N_y=80\sim200$ uniform cells in both the $x$ and $y$ directions.
The number of discretized cells in the $x$ or $y$ direction depends on the spatial period $P$.
The larger the spatial period $P$ is, the more discretized cells are used.

The phonon dispersion and scattering in silicon and germanium materials are given in Appendix~\ref{sec:dispersionscattering}.
For each phonon branches (LA/TA), the wave vector is discretized into $N_B$ equally and the mid-point rule is used for the numerical integration of the frequency space. In total, $2N_B$ discretized frequency bands are considered.
Here we set $N_B =20$.

The three-dimensional solid angle is $\bm{s}=\left( \cos \theta, \sin \theta \cos \varphi, \sin \theta  \sin \varphi  \right)$, where $\theta \in [0,\pi]$ is the polar angle and $\varphi \in [0,2\pi]$ is the azimuthal angle.
The $\cos \theta \in [-1,1]$ is discretized with the $N_{\theta}$-point Gauss-Legendre quadrature, while the azimuthal angular space $\varphi \in [0,\pi]$ (due to symmetry) is discretized with the $\frac{N_{\varphi}}{2}$-point Gauss-Legendre quadrature.
In this study, we set $N_{\theta} \times N_{\varphi} =40 \times 40$.

The van Leer limiter is used to deal with the spatial gradient of the distribution function and the time step is
\begin{align}
\Delta t= \text{CFL} \times \frac{\Delta x}{v_{\text{max}} },
\label{eq:CFLnumber}
\end{align}
where $\Delta x$ is the minimum discretized cell size, $\text{CFL}$ is the Courant–Friedrichs–Lewy number and $v_{\text{max}} $ is the maximum group velocity.
In this simulations, $\text{CFL}=0.4$.

The isothermal boundary condition is used for the heat sink, where the incident phonons are all absorbed and the phonons emitted from the boundary are the equilibrium state with the boundary temperature $T_{BC}$.
Its mathematical formula is
\begin{align}
e(T_{BC}, \bm{s}, \omega ) = C (T_{BC} - T_0), \quad   \bm{s} \cdot  \mathbf{n} >0,
\label{eq:isothermal}
\end{align}
where $\mathbf{n}$ is the normal unit vector of the boundary pointing to the computational domain.
The diffusely reflecting adiabatic boundary condition controls the total heat flux across the boundary to be zero and phonons with the same frequency reflected from the boundary are equal along each direction.
Its mathematical formula is
\begin{align}
e( \bm{s}, \omega ) = C ( T_{w}- T_0), \quad   \bm{s} \cdot  \mathbf{n} >0,
\label{eq:diffusely}
\end{align}
where
\begin{align}
T_{w}   = T_0 +  \frac{ - \sum_p \int \int_{ \bm{s}' \cdot \mathbf{n} <0 }   v_g e  \bm{s}' \cdot \mathbf{n} d\Omega d\omega }{ \sum_p \int \int_{\bm{s} \cdot \mathbf{n} >0 } v_g C  \bm{s} \cdot \mathbf{n} d\Omega d\omega } .
\end{align}

{\color{blue}{
\section{Independence tests of the discrete parameters}
\label{sec:validation}

Independence tests of the discrete parameters in the whole phase space are conducted.
Firstly, the bulk thermal conductivity $\kappa_{bulk}=\sum_p \int  C v_{g}^{2} \tau /3 d\omega  $ is used to find the optimized number of frequency bands.
With different discretized numbers of the phonon frequency bands $N_B =20,~ 40,~ 100$, the calculated bulk thermal conductivities of silicon at room temperature are all $145.9$ W/(m$\cdot$K).
Similarly, the bulk thermal conductivity of germanium at room temperature is $58.8$ W/(m$\cdot$K).
Hence the numerical integration in the phonon frequency space is regarded as converged when $N_B \geq 20$.

Secondly, the independence tests of the discretized solid angle space and time step are implemented, which is necessary to ensure that the ray effect and false scattering has little effect on the numerical results.
We take the quasi-2D hotspot system (\cref{close_packed2D}) as an example.
The ray effects usually appear in the ballistic regime, so that we simulate the case with $\text{Kn}=10$ and $P^* =1$.
The temporal evolution processes of the average temperature with different discretized solid angles and CFL numbers are plotted in~\cref{Independence_tests}.
The numerical results confirm that the choice of $N_{\theta} \times N_{\varphi}=40 \times 40$ and $\text{CFL}=0.4$ are enough to accurately predict the transient ballistic heat conduction.

Finally, the discretized spatial cells are also tested.
Usually, more discretized cells are needed near/in the diffusive regime, so that we simulate the case with $\text{Kn}=0.1$ and $P^* =4$.
Different discretized numbers in the $x$ direction is tested and it can be found that $80$ discretized cells in the $x$ direction and $\text{CFL}=0.4$ are adequate.

In summary, the present discretizations in Appendix~\ref{sec:solver} are accurate to capture the multiscale transient heat conduction in 3D materials.
}}

\bibliography{phonon}

\end{document}